\documentclass[10pt,%
			   amsmath,amssymb,%
			   reprint,
			   floatfix,
			   aip]{revtex4-2}

\usepackage[utf8]{inputenc}
\usepackage[T1]{fontenc}

\usepackage{graphicx,algorithmic}

\usepackage{newfloat}

\DeclareFloatingEnvironment[
    fileext=loa,
    listname=List of Algorithms,
    name=ALGORITHM,
    placement=tbhp,
]{algorithm}

\usepackage{multirow,lipsum,xspace}
\usepackage[usenames,dvipsnames]{xcolor}

\definecolor{red}{RGB}{146,19,11}
\definecolor{plum}{RGB}{71,61,107}
\definecolor{blue}{RGB}{10,85,145}
\definecolor{gray}{RGB}{95,95,95}
\definecolor{black}{RGB}{0,0,0}
\usepackage{siunitx}

\newcommand{\cmp}[1]{O(#1)}

\newcommand{\avg}[1]{\langle#1\rangle}

\newcommand{\argmax}[0]{\arg\!\max}
\newcommand{\argsmax}[0]{{\rm args}\!\max}
\newcommand{\random}[0]{{\rm random}}
\newcommand{\shuffle}[0]{{\bf shuffle}\,}
\newcommand{\pushs}[0]{{\bf push}\,}
\newcommand{\pops}[0]{{\bf pop}\,}

\newcommand{\lpa}[0]{LPA\xspace}
\newcommand{\flpa}[0]{FLPA\xspace}

\newcommand{\figref}[1]{Fig.~\ref{fig:#1}\xspace}
\newcommand{\tblref}[1]{Table~\ref{tbl:#1}\xspace}
\renewcommand{\eqref}[1]{Eq.~(\ref{eq:#1})\xspace}

\graphicspath{ {figures/} }
\DeclareGraphicsExtensions{.pdf,.eps,.png,.jpg}

\begin{document}

\title{Large network community detection by fast label propagation}

\author{Vincent A.\ Traag}
\email{v.a.traag@cwts.leidenuniv.nl}
\affiliation{Leiden University, Centre for Science and Technology Studies, Leiden, The Netherlands}

\author{Lovro Šubelj}
\affiliation{University of Ljubljana, Faculty of Computer and Information Science, Ljubljana, Slovenia}


\begin{abstract}
Many networks exhibit some community structure. There exists a wide variety of approaches to detect communities in networks, each offering different interpretations and associated algorithms. For large networks, there is the additional requirement of speed. In this context, the so-called label propagation algorithm (\lpa) was proposed, which runs in near-linear time. In partitions uncovered by \lpa, each node is ensured to have most links to its assigned community. We here propose a fast variant of \lpa (\flpa) that is based on processing a queue of nodes whose neighbourhood recently changed. We test \flpa exhaustively on benchmark networks and empirical networks, finding that it can run up to $700$ times faster than \lpa. In partitions found by \flpa, we prove that each node is again guaranteed to have most links to its assigned community. Our results show that \flpa is generally preferable to \lpa.
\end{abstract}

\flushbottom
\maketitle
\thispagestyle{empty}

\section*{Introduction}

\noindent
Networks are relevant in various scientific fields, ranging from social networks in sociology to metabolical networks in biology.
There are various techniques to try to improve our understanding of networks.
One such technique is to cluster the nodes of a network, such that nodes within a cluster are relatively densely connected while they are relatively sparsely connected between clusters.
There is a wide variety of clustering approaches to networks, such as modularity\cite{NG04} and stochastic block models\cite{Pei20} and approaches based on dynamical processes on networks\cite{RB07}, such as random walks\cite{RB08}.
Sometimes, a similar approach can be solved in various ways.
For example, when modularity was first proposed, it used a cutting approach based on betweenness\cite{NG04}.
New algorithms were continuously proposed, either improving the speed of the algorithm or the quality of the partition.
This includes a fast greedy approach\cite{Clauset2004-bf}, a slower simulated annealing approach\cite{Reichardt2006-ij}, a faster algorithm based on extremal optimisation\cite{Duch2005-jj}, a fast hierarchical multi-level method, known as the Louvain algorithm\cite{BGLL08} which was most recently improved upon in the Leiden algorithm\cite{TWV19}.
Most of these algorithm can also optimise different quality functions, such as the Constant Potts Model\cite{Traag2011-nu}.

A technique that takes a heuristic approach is the label propagation algorithm (LPA)\cite{RAK07}.
Its foremost focus is on speed, trying to find clusters in as little time as possible.
Simply put, LPA works by iteratively updating the label of each node to a label that is most common among its neighbours.
We here propose a fast variant of LPA (FLPA), which can potentially run up to hundreds of times faster than LPA.
This allows to cluster even larger networks in even less time.
We consider this to be useful as a first initial look at a network, although other methods are arguably more robust and preferable\cite{Sub20}.
The results of LPA are only local minima of a global quality function for which the optimum is simply placing all nodes into a single community\cite{Tibely2008-qd}.
Other quality functions may be more informative of any structure in the network.

We first briefly review LPA and introduce the fast variant FLPA in the next section.
We then briefly analyse the performance of LPA and FLPA theoretically, followed by experimental analyses on both synthetic and empirical networks.

\section*{Label propagation algorithm}

\noindent
We now introduce the label propagation algorithm (LPA)\cite{RAK07} more formally.
For a more detailed review of label propagation algorithms, we refer the reader to literature reviews\cite{Sub20,Garza2019-td}.

Let $G = (V, E)$ be an undirected multigraph with nodes $V$ and edges $E$, where there are $n = |V|$ nodes and $m = |E|$ edges.
Let $A$ be the adjacency matrix of graph $G$, such that $A_{ij}$ is the number of edges between $i$ and $j$, with $A_{ij} = 0$ if and only if nodes $i$ and $j$ are not connected (i.e.\ $(i, j) \notin E$).
The implementation of LPA is quite straightforward.
Let $c_i$ be the label of node $i \in V$.
Typically, each node is initially labelled differently, i.e. $c_i = i$ for all $i \in V$.
At each step, we take a random $i \in V$ and change its label to the majority in its neighbourhood.
In more detail, we do the following.
For a specific node $i$, we count how many neighbours have label $c$ as $n_c = \sum_{j\in V} A_{ij}\delta(c_j, c)$, where $\delta(c_j, c)$ is the Kronecker delta function such that $\delta(c_j, c) = 1$ if $c_j = c$ and $0$ otherwise.
We then consider the set of most frequent labels $\{c\} = \argmax_c n_c$.
We randomly sample uniformly from the set of most frequent labels $\{c\}$ a label $c^*$ and update the label $c_i = c^*$.
We repeat these steps over all nodes in $V$.
After having looped over all nodes, we check whether all labels are among most the frequent, whether the label is \emph{maximal}.
If any label is not maximal, we perform another iteration over all nodes, until all labels are maximal.
This algorithm is summarised in Algorithm~\ref{alg:lpa}.

This version of the label propagation is also referred to as the asynchronous implementation of label propagation\cite{RAK07}.
The synchronous implementation of label propagation showed potential problems with (near) bipartite networks and some other networks, resulting in oscillations of labels.
We therefore do not consider the synchronous implementation, and limit the discussion to the asynchronous implementation.

After LPA terminates, it is guaranteed that the label $c_i$ of each node $i\in V$ is maximal.
That is, $n_{c_i} = \max_c \sum_{j\in V} A_{ij}\delta(c_j, c)$.
This is trivial to prove, since LPA continues to process nodes until all labels are maximal.
In the original introduction of LPA\cite{RAK07}, they observed that this is close to the definition of ``communities in the strong sense'' as introduced by \citet{Radicchi2004-yn}.

The overall time complexity of a single iteration over all nodes in LPA is $\cmp{m}$.
The number of iterations necessary before convergence is not know theoretically.
It was observed that generally only a few iterations suffice to have most labels consistent with their final labelling\cite{RAK07}.
However, there are no clear results for the overall runtime complexity of LPA.

\subsection*{Retention strategy}

\noindent
LPA stops iterating whenever the labels of all nodes are maximal.
The original implementation of LPA\cite{RAK07} simply considered always updating the label with a randomly selected maximal label.
Later, a so-called retention strategy was suggested: update the label only if it is not maximal\cite{Barber2009-vb}.
This has one great benefit: we can simply keep track of whether a label was updated during the iteration, and if so, we continue iterating over all nodes.
This means it is not necessary to check for maximal labels after an iteration over all nodes, making the implementation more efficient.
The retention strategy is summarised in Algorithm~\ref{alg:lpar}.

In addition, this retention strategy introduces more stability, because it does not continuously sample from competing maximal labels for a single node like the original LPA does.
LPA sometimes shows the appearance of a ``giant'' cluster\cite{RAK07,LHLC09} which might be related to the quality function which it implicitly optimises\cite{Tibely2008-qd}.
Indeed, merging neighbouring clusters does not alter the label maximality, so that any arbitrary combination of clusters in principle will still meet the original stopping criteria.
The retention strategy might prevent the method from finding such ``giant'' clusters\cite{Barber2009-vb}.

Some empirical observations in the literature noted that we can expect roughly $1.03 m^{0.23}$ iterations for the retention strategy on empirical networks\cite{Subelj2011-cb}.
This leads to an overall time complexity of about $O(m^{1.23})$ for the retention strategy.

\subsection*{Fast label propagation algorithm}

\noindent
We now introduce the fast label propagation algorithm (FLPA).
It is based on the same principle that is used for the fast local move in the Leiden algorithm\cite{TWV19}.
Similar to LPA, each node $i \in V$ has an associated label $c_i$, and we use a similar majority update rule.
However, instead of checking after an iteration whether all labels are maximal, or by considering whether labels are updated, as done in the retention strategy, we maintain an explicit queue of nodes $Q$ that should be considered.
If $c_i$ of node $i$ is changed, we append some of its neighbours $N_i = \{ j \mid (i, j) \in E \}$ to the queue.
In particular, we add each neighbour $j \in N_i$ to the queue that has a label different from the \emph{new} label of $i$, $c_j \neq c_i$ and does not yet belong to the queue $j \not\in Q$.
At each step, we pop the node from the beginning of the queue, and we continue to process all nodes until the queue is empty.
Hence, instead of iterating over \emph{all} nodes if a label is changed, we only consider nodes in whose neighbourhood a label changed.
This greatly reduces the number of nodes that we consider, making the algorithm even faster.
The algorithm is summarised in Algorithm~\ref{alg:flpa}.

We now prove that FLPA provides the same guarantee as LPA, namely that after FLPA terminates, it is guaranteed that the label $c_i$ of each node $i\in V$ is maximal.
We first observe that a node $i$ that is not in the queue will have its label $c_i$ as its maximal label.
That is, $n_{c_i} = \max_c \sum_{j \in V} A_{ij}\delta(c_j, c)$.
This clearly holds when node $i$ was processed.
If node $i$ is currently not part of the queue, we can discern two cases.
In the first case, no labels in its neighbourhood have changed at all.
In this case, the label $c_i$ continues to be the maximal label.
In the second case, the label of a neighbour $j\in N_i$ changed.
It must then hold that $c_j = c_i$, since otherwise $i$ should have been added to the queue.
In that case, the number of labels in the neighbourhood of $i$ that equal $c_i$ increased.
If $n_{c_i}$ was maximal prior to node $j$ changing its label, it continues to be maximal.
Hence, $c_i$ continues to be the maximal label as long as node $i$ is not part of the queue.
If the queue is empty, the label $c_i$ of each node $i$ is guaranteed to be the maximal label, similar to LPA.

Earlier literature also suggested a speedup of LPA in a similar fashion\cite{XS11,TASGIN2019315}.
However, they seem to have taken a slightly more complicated approach, either introducing additional heuristics or requiring the algorithm to check whether a neighbour might be updated or not.
Our approach is easier to implement, and seems to result in even greater speedups.

\section*{Results}

To understand better the differences between LPA, its retention alternative, and FLPA, we analysed three theoretical graphs: a complete graph, a star graph and a cycle graph.
In addition, we also analysed the differences in results in practice.
We ran benchmarks on five different types of synthetic networks and on twelve different empirical networks.
In addition to comparing LPA, its retention alternative, and FLPA, we also compare with the Leiden algorithm, which is one of the fastest available algorithms\cite{TWV19}.
We use the Leiden algorithm to optimise for modularity.
We compare both the speed and the resulting partitions.

We first discuss our theoretical results.
We then present the results for the synthetic networks\cite{newman_networks_2018}.
Following that, we discuss the results for the empirical networks.

\subsection*{Theoretical analysis}

LPA performs a number of iterations over $n$ nodes.
Each potential update of a label for a node $i$ has a complexity of $\Theta(k_i)$, where $k_i$ is the degree of node $i$, and therefore the total complexity of a single iteration is linear in the number of edges $\Theta(m)$.
If the number of iterations does not scale with $n$, we simply have linear complexity $\Theta(m)$.
Presumably, however, the number of iterations will increase with $n$, but it is not clear exactly how.
The complexity also relates to the resulting partitions, since finding a partition of $n/2$ equally weighted clusters will most likely take less time than finding a partition consisting of a single cluster.
The same reasoning applies to the retention variant and FLPA.

For some specific graphs we can analyse the complexity in more detail.
We analyse the algorithms for three theoretical graphs: a complete graph, a star graph and a cycle graph.
We summarise the runtime complexities in \tblref{complexity}.

\begin{table}
    \centering
	\begin{tabular}{|cccc|}\hline
		Graph & LPA & retention & FLPA \\\hline
		  \multirow[b]{3}{*}{\includegraphics[height=3em]{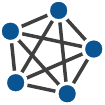}} & \multirow{3}{*}{$\Omega(2m)$} & \multirow{3}{*}{$\Theta(2m)$} & \multirow{3}{*}{$\Theta\left(m + \frac{1}{n - 2}\right)$} \\
	       & &  &  \\
            & & & \\\hline
        \multirow[b]{3}{*}{~\includegraphics[height=3em]{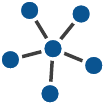}~} 
        & \multirow{3}{*}{$\Theta\left(2m + \frac{2n(n - 2)}{(n - 1)^2}\right)$}
        % & \multirow{3}{*}{$\Theta\left(2n\left[1 + \frac{n - 2}{(n - 1)^2}\right]\right)$}
        & \multirow{3}{*}{$\Theta(2m)$} 
        & \multirow{3}{*}{$\Theta\left(m + \frac{n - 2}{(n - 1)^2}\right)$} \\
	       & &  & \\
             & & & \\\hline
	\end{tabular}
	\caption{Expected runtime complexity of LPA, its retention variant and FLPA for two theoretical graphs.}
	\label{tbl:complexity}
\end{table}

\subsubsection*{Complete graph}
Let us start by analysing a complete graph with $n$ nodes and $m = \binom{n}{2}$ edges. % {n \choose 2}
All three algorithms, LPA, its retention variant, and FLPA, will find a partition consisting of a single cluster within a single iteration of all nodes.
Suppose on the contrary that a partition consists of $k > 1$ clusters, while the label of each node is maximal.
We prove by contradiction that this is not possible.
With $k$ clusters, there are $k$ labels, and each label $c$ occurs $n_c$ times.
Now suppose that $n_c \geq n_d$ for a pair $c \neq d$.
Then each node with label $d$ has only $n_d - 1$ neighbours with label $d$, and since $n_c > n_d - 1$ its own label $d$ is not maximal.
Therefore $n_c < n_d$ for all $c$ and $d$, which is impossible, and there can be only one cluster.

Let us now consider the complexity.
Initially, each algorithm starts with a singleton partition such that $c_i = i$.
Let us analyse the first node that is considered in each algorithm.
Without loss of generality, we can label this node $1$.
Each node has $n - 1$ neighbours, and initially each node is in its own cluster, meaning there are $n - 1$ unique labels for node $1$, each of which occurs only once.
Since the current label $c_1 = 1$ of node $1$ is not yet maximal, a random label will be chosen by each algorithm, say label $i$, and we set $c_1 = i$.
Then, when we consider node $2$ (assuming $i \neq 2$), there are $n - 2$ unique labels, of which $n - 3$ occur once and one label (namely label $i$) occurs twice.
Thus, there is only a single maximal label $i$, and node $2$ will switch to label $i$.
Subsequently, all remaining nodes will also switch to label $i$, and hence all nodes will be assigned label $i$ in the end.

All three algorithms consider $n$ nodes in this case.
However, both LPA and its retention variant will consider again $n$ nodes for updating, or checking for maximality.
FLPA does not need to do this, and the queue will be empty after considering all $n$ nodes.
This means that the total runtime in this case is $\Theta(2m)$ for LPA and retention, and $\Theta(m)$ for FLPA.

There is an exception, namely if $i = 2$, which occurs with probability $\frac{1}{n - 1}$.
Let us consider what happens to node $2$.
If $i = 2$, its current label $c_2 = 2$ is maximal, since it has $n - 1$ unique labels, each of which occurs only once, but now this includes its own label $2$.
The retention strategy will not draw a random label, since its current label is already among the maximal labels of its neighbours.
Therefore, for retention, all remaining nodes will simply switch to label $2$ and we end up with a complexity of $\Theta(2m)$.
However, in LPA and FLPA, a random label is drawn from all maximal labels.
With probability $\frac{1}{n - 1}$, label $2$ is drawn. In this case, all other remaining labels also change to label $2$ and all labels are maximal after a single iteration.
With probability $1 - \frac{1}{n - 1}$, label $j \neq 2$ is drawn.
In this case, we need to perform another round in LPA since $c_1 = 2$ is not maximal, and in FLPA node $1$ is added to the queue again.

The probability of needing a second iteration over all nodes for LPA is then
\begin{equation}
	 \frac{1}{n - 1} \left(1 - \frac{1}{n - 1}\right),
    \nonumber
\end{equation}
which goes to $0$ for $n \to \infty$.
This covers only the first iteration, and there will be similar probabilities involved in each subsequent iteration, but clearly those probabilities will equally go to $0$.
Therefore, the expected runtime for LPA is close to $\Omega(2m)$, although slightly higher.

If nodes continuously select the label of the node that will be considered next, as in $c_1 = c_2$, there will continue to be a single label that occurs twice, while all other labels occur once.
In FLPA, if a node chooses the label of the node that will be considered next, i.e. $c_i = c_{i + 1}$ for $i>1$, we add the node $i-1$ that was previously considered to the queue.
For each node, this happens with probability $\frac{1}{n - 1}$, which leads to an expected number of additional nodes of
\begin{equation}
	 \frac{1}{n - 1} \sum_{k = 1}^{\infty} \left(\frac{1}{n - 1}\right)^k = \frac{1}{(n-1)(n-2)}. 
  \nonumber
\end{equation}
Since the degree of each node is equal to $n-1$, the total expected runtime for FLPA is $\Theta\left(m + \frac{1}{n - 2}\right)$.

\subsubsection*{Star graph}

We now analyse a star graph of $n$ nodes, with a single node in the center and $n - 1$ leaf nodes connected to the central node, with in total $m = n - 1$ edges.
Similar to the complete graph, all three algorithms will find a partition consisting of a single cluster.
Clearly, the leaf nodes cannot have a different label as the central node, since that is their only neighbour.
So, by definition, all leaf nodes must have the same label as the central node, and hence all nodes have the same label.

Regardless of the updating order, all leaf nodes will always adopt the label from the central node in each algorithm.
The only question is what happens for the central node.
If the central node is selected first, it will simply choose a random label from the leaves, and all leaves will adopt that label.
If the central node is selected third or later, it will not change its label because its own label is then by definition the only maximal label.
Thus, in both these cases, only $n$ nodes are considered, and both LPA and retention perform another pass over all nodes to check for maximality (in the case of LPA) or because there was a change (in retention), resulting in a total runtime of $\Theta(2m)$.
In FLPA, no nodes are ever added to the initial queue of $n$ nodes, resulting in a runtime of $\Theta(m)$.

Let us examine what happens when the central node is selected as the second node to update.
This only happens with probability $\frac{1}{n - 1}$ and is therefore unlikely to have much effect on the overall expected runtime.
For the retention strategy, if the central node is considered second, this means that the first node has already adopted the label of the central node and the retention strategy will not update its label.
Thus, retention is not affected by this and maintains a runtime of $\Theta(2m)$.

In LPA, if the central node is considered second, it will choose the same label as its current label with probability $\frac{1}{n - 1}$.
In this case, after all the remaining nodes have also updated their label, all nodes have identical labels and LPA terminates.
If a different label is selected, all other $n - 2$ leaf nodes also adopt that label.
In a subsequent iteration, no node except the first leaf node initially considered will update its label.
Therefore, LPA will converge in at most two iterations, with the second iteration occurring with probability $\frac{1}{n - 1}\frac{n - 2}{n - 1}$, resulting in a total expected runtime of
\begin{equation}
	 \Theta\left(2m\left[1 + \frac{n - 2}{(n - 1)^2}\right]\right). \nonumber
\end{equation}

Finally, in FLPA, if the central node is considered second and it chooses a label different from its current label, the first leaf node initially considered will be added to the queue, and nothing more.
Hence, the total expected runtime is
\begin{equation}
	\Theta\left(m + \frac{n - 2}{(n - 1)^2}\right). \nonumber
\end{equation}

\subsubsection*{Cycle graph}

Let us now consider a cycle graph of $n$ nodes. In contrast to the complete graph and the star graph, each algorithm now results in different partitions.
In all three algorithms, each node whose neighbours have not yet been updated, simply chooses the label of a random neighbour.
The distribution of cluster sizes is difficult to analyse exactly.
Regardless, all labels are maximal only if all clusters are larger than a single node.
This is clear for the nodes in the interior of a cluster, since their label is identical to that of their two neighbours.
For the nodes at the border of a cluster, their current label is among the maximal labels.
At the borders of clusters is where differences between the individual algorithms emerge.

For the retention strategy, no updates will be considered anymore.
Therefore, it will quickly settle on any partition that does not include a cluster of size one.

For LPA the picture is a bit different.
At each border of two clusters, the node updates its label by randomly choosing between the label on its left and the label on its right.
Thus, as long as there is still a cluster of size one in the partition, all borders will continue to change in LPA, unlike for the retention strategy.
Although clusters of size one may disappear, they may also newly appear when clusters of larger size shrink.

For FLPA the picture is again slightly different.
Whereas LPA simply continues to move around all nodes, and thus moves around all borders as long as there is a cluster of size one, FLPA has more local dynamics.
Suppose that FLPA updates the label of a node at the border of a cluster.
If this happens, its neighbour in the old cluster is added to the queue (unlike its neighbour in the new cluster).
So with probability $\frac{1}{2}$ a neighbouring node will be updated, and this happens again with probability $\frac{1}{2}$ et cetera.
Continuing this reasoning, the expected number of moves resulting from a single border is 
\begin{equation}
	  \sum_{k = 0}^\infty \left(\frac{1}{2}\right)^k = 2. \nonumber
\end{equation}
Finally, a node in a cluster of size one is never maximal, so FLPA will continue to run until there are no more such clusters.
The difference, however, is that only nodes in such clusters can be moved, rather than always considering all nodes at the borders like LPA does.

In summary, the retention strategy will most likely detect the most fine-grained clusters, LPA the least fine-grained clusters, and FLPA somewhere in between.
Experimental simulations suggest that retention settles on clusters with $2.72$ nodes and FLPA on clusters with $4.11$ nodes, whereas
the cluster size increases with the number of nodes $n$ for LPA.
%The exact runtime is also more challenging to analyse.
In terms of the runtime, retention is then expected to converge the fastest, LPA the slowest, and FLPA somewhere in between.
Nonetheless, retention will still consider updating all $n$ nodes, even when moving only a single node in a cluster of size one.
So, in total, the runtime of FLPA might still be lower than the runtime of the retention strategy.

\subsection*{Experimental results}

We now present some experimental results based on implementations of LPA, its retention variant, FLPA, and the Leiden algorithm for optimising modularity.
For all test results, we report averages and standard deviations of $2\,000$ runs unless explicitly stated otherwise.
All results were run on Dell PowerEdge M620 computing nodes with Intel E5-2697 CPUs.

\begin{figure*}
	\centering
    \includegraphics[width=\textwidth]{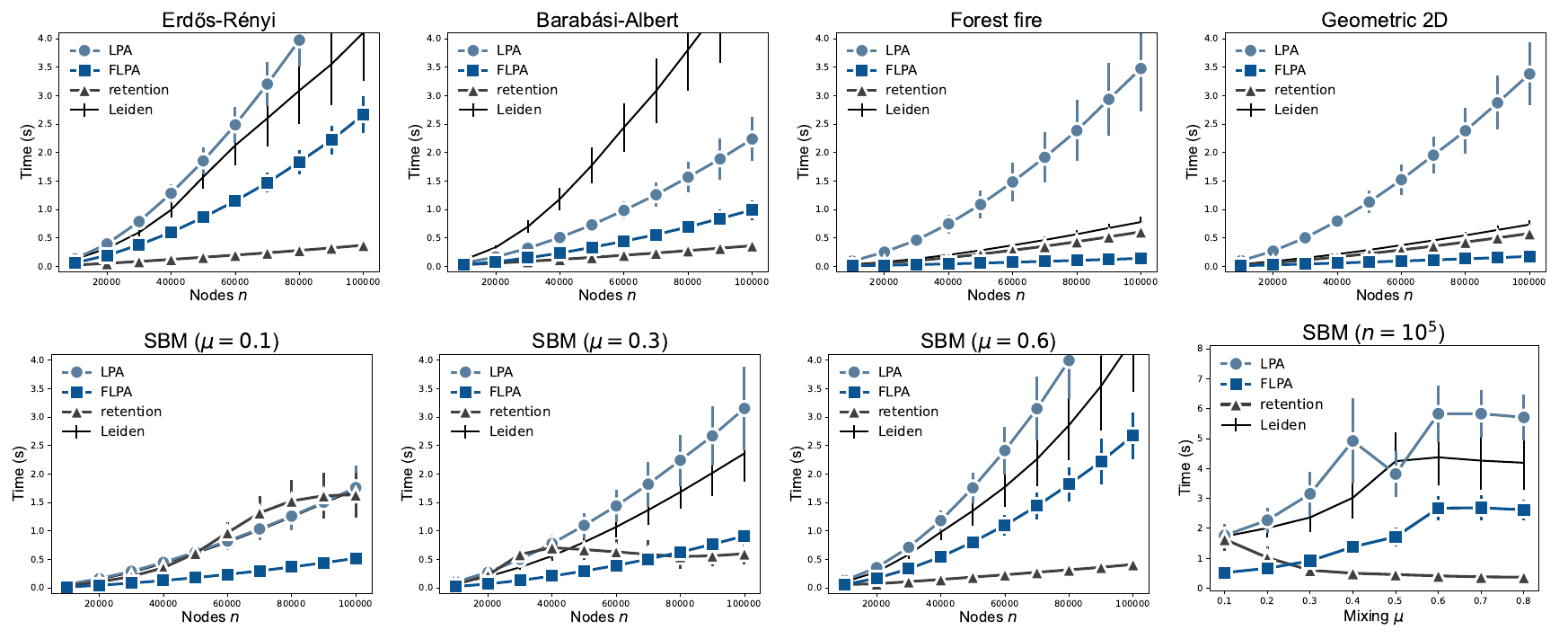}
	\caption{Algorithm runtime for synthetic networks with increasing number of nodes $n$ and  the average degree $\avg{k}=10$. 
	(top) Erdős-Rényi random graphs, Barabási-Albert scale-free graphs, forest fire graphs with burning probability $0.5$ and two-dimensional geometric graphs with connection radius $\sqrt{\avg{k}/(\pi n)}$. 
	(bottom) Stochastic block model graphs with $100$ groups and mixing parameter $\mu$. 
	%Results are averages over $2\,000$ runs of the algorithms and error bars show standard deviations.
 }
	\label{fig:time}
\end{figure*}

\subsubsection*{Synthetic networks}

As is clear, for all networks FLPA is always faster than LPA (Fig.~\ref{fig:time}).
For the largest networks with $100\,000$ nodes, FLPA is somewhere between $3$--$10$ times faster than LPA.
On Erdős-Rényi (ER) graphs, LPA is only about twice as slow as FLPA, which is comparable to the runtime of the Leiden algorithm.
On two-dimensional geometric graphs, LPA is even $3$--$4$ times slower than the Leiden algorithm, which in turn is still about $3$--$4$ times as slow as FLPA.
We tested the algorithms on a stochastic block model (SBM) of $100$ groups and a mixing parameter of $\mu=0.6$ with average degree of $\avg{k}=10$, such that each node has about $4$ links within its own group and $6$ scattered across the other groups.
For this case, LPA is again about twice as slow as FLPA, which is again about $1.5$ times faster than the Leiden algorithm.

\begin{figure}
	\centering
    \includegraphics[width=0.5\textwidth]{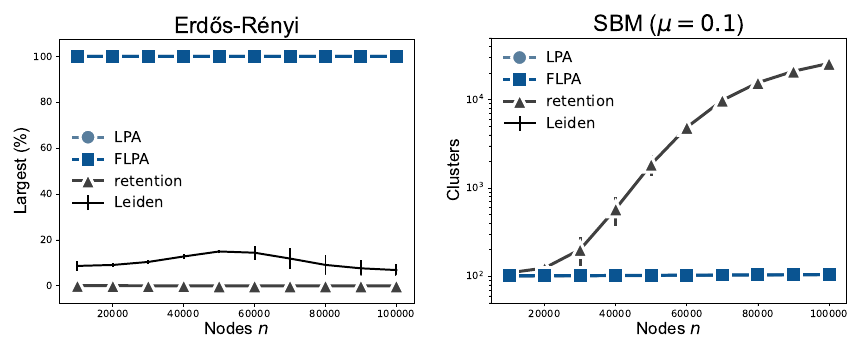}
	\caption{(left) Largest cluster size for Erdős-Rényi random graphs and (right) the number of clusters for stochastic block model graphs 
    with increasing number of nodes $n$ and the average degree $\avg{k}=10$.
    % with %$100$ groups and 
    %mixing parameter $\mu=0.1$. 
	%Results are averages over $2\,000$ runs of the algorithms and error bars show standard deviations.
        Note that the results for LPA are not well visible because they overlap with the results of FLPA.}
	\label{fig:stats}
\end{figure}

The comparison between FLPA and the retention variant of LPA is more complex.
For some networks, the retention strategy alternative is faster, while for other networks FLPA is faster.
However, in cases when the retention strategy is faster, it typically fails to perform well.
For instance, in ER graphs, both LPA and FLPA find a single large cluster (Fig.~\ref{fig:stats} left), as expected based on earlier literature\cite{RAK07}, but the retention strategy simply terminates almost immediately, finding a partition that closely resembles the singleton partition of each node in its own cluster.
In SBM graphs, both LPA and FLPA typically find partitions that are close to the planted partition, while the retention strategy finds many small clusters within each group of the planted partition (Fig.~\ref{fig:stats} right).
This is understandable, since every group is essentially an ER graph internally.
In short, the retention strategy is faster simply because it stops very soon after having found some very small community structure.

The overall scaling of both LPA and FLPA seems to be near linear, although there might be some super-linear factors.
For all algorithms, we experimentally estimate runtime complexities of the form $a m^b + c$ using the Levenberg-Marquardt algorithm, where $m$ is the number of edges and the coefficient $b$ is of central interest.
For ER graphs, LPA scales as $O(m^{1.58})$ and FLPA as $O(m^{1.49})$, while the retention strategy scales as $O(m^{1.15})$.
As we already noticed, in ER graphs, retention finds very small clusters, which explains its lower complexity.
On forest fire graphs and geometric graphs, FLPA is much faster than LPA and the retention strategy.
Indeed, we find a scaling of $O(m^{1.27})$ for forest fire graphs and $O(m^{1.21})$ for geometric graphs for FLPA, while LPA shows a scaling of $O(m^{1.67})$ and $O(m^{1.57})$ respectively, with retention showing a scaling of $O(m^{1.47})$ and $O(m^{1.37})$ respectively.
Finally, on SBM graphs, the performance of LPA and FLPA is similar to the performance on ER graphs, leading to a complexity of $O(m^{1.53})$ for LPA and $O(m^{1.64})$ for FLPA.
For the retention strategy, the scaling cannot be estimated unambiguously, since it finds completely different partitions depending on the size of the graph.
That is, for larger graphs, it tends to find more fine-grained structure within each cluster, causing it to converge relatively faster than for smaller graphs.

\begin{figure*}[b]
	\centering
    \includegraphics[width=0.75\textwidth]{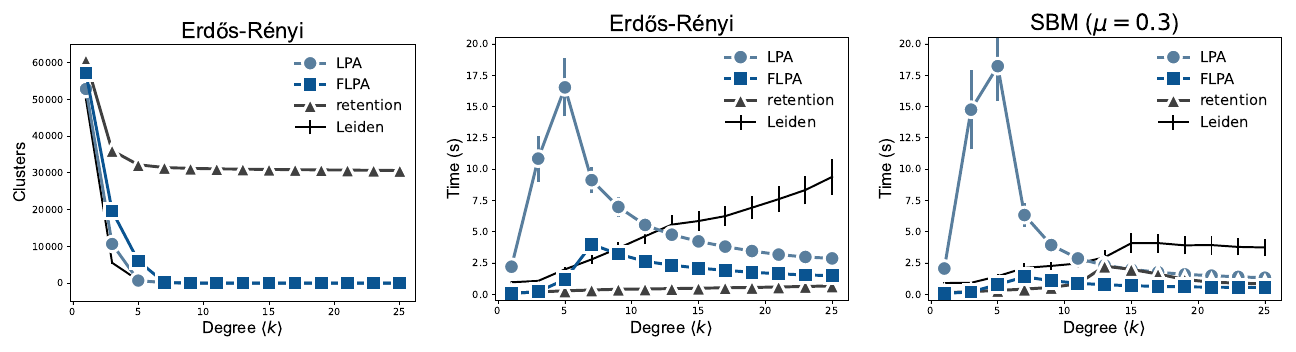}
	\caption{Algorithm runtime for (middle) Erdős-Rényi random graphs and (right) stochastic block model graphs with $n=10^5$ nodes and varying average degree $\avg{k}$.}
	\label{fig:time_degree}
\end{figure*}

The runtime complexity also depends on how challenging it is to find a partition. % the type of graph, and
To investigate this more closely, we consider the runtime for each algorithm while varying the average degree $\avg{k}$ (\figref{time_degree}).
LPA, FLPA and retention all find small clusters in ER graphs with low average degree.
If the degree is sufficiently large, both LPA and FLPA will find a single large cluster (\figref{time_degree} left), as also suggested by our theoretical analysis of a complete graph.
The larger the degree, the faster both LPA and FLPA converge towards this large cluster (\figref{time_degree} middle).
Ultimately, when $\avg{k} = n-1$, the graph is a complete graph, for which our theoretical runtimes indicate that FLPA is about twice as fast as LPA.
In contrast, the retention strategy does not show a convergence to a single large cluster for $\avg{k} < 25$.

%\redo{The retention strategy converges much more slowly to this single large cluster. As the degree increases, retention finds larger and larger clusters, so the running time increases with the degree.}
% The retention strategy converges much slower to this single large cluster. 
% %, but will eventually also converge to a single cluster in ER networks.
% With larger degree, retention finds increasingly larger clusters, as a result of which the runtime increases with increasing degree.

The overall figure is very similar for SBM graphs (\figref{time_degree} right).
Initially, all algorithms struggle to uncover the planted partition. 
When the degree is sufficiently high, the planted partition becomes more easily recognisable, and the algorithms converge faster.

% Initially, with low degree, all algorithms struggle to uncover the planted partition, which will take quite some time. 
% When the degree is sufficiently high, the planted partition becomes more easily discernible, and the algorithms converge more quickly.
%Increasing the degree further seems to affect little the runtime complexity, presumably because the convergence to the planted partition is sped up.

\begin{figure*}
	\centering
    \includegraphics[width=0.85\textwidth]{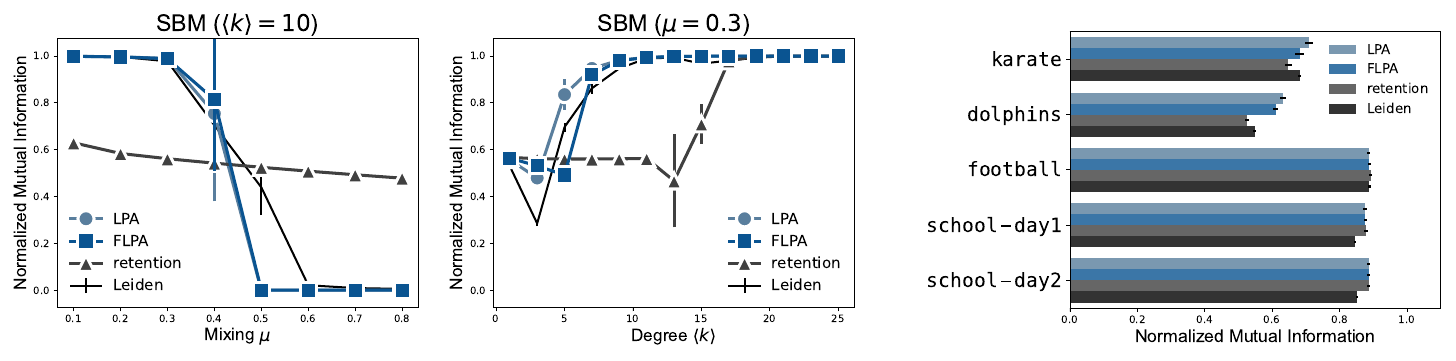}
	\caption{
    Accuracy of partitions for
    stochastic block model graphs with $n=10^5$ nodes and (left) varying mixing parameter $\mu$, where higher $\mu$ corresponds to partitions that are more challenging to detect, or (middle) varying average degree $\avg{k}$, and (right) small social networks with the metadata on node clusters.}
	\label{fig:nmi_SBM_soc}
\end{figure*}

For SBM graphs, we also compare the detected partitions with the planted partition of $100$ groups (\figref{nmi_SBM_soc}) using the normalised mutual information (NMI)\cite{danon_comparing_2005}.
When we increase the mixing parameter $\mu$ (\figref{nmi_SBM_soc} left), we find that both LPA and FLPA are able to detect the correct partition up to a threshold.
For $\mu = 0$ all edges are within groups, while for $\mu=1$ all edges are between groups, and so increasing $\mu$ makes it more difficult to correctly detect the planted partition.
For $\mu = \frac{n - n_c}{n - 1}$ the SBM is identical to an ER graph (where $n_c$ is the size of the communities), although it already becomes essentially indistinguishable before this threshold due to stochastic fluctuations\cite{Floretta2013-zh}.
Up to $\mu =0.3$ both LPA and FLPA detect the planted partition perfectly, for $\mu = 0.4$ the performance starts to degrade, and for $\mu = 0.5$ the algorithms no longer find the planted partition at all.
This closely resembles the results of the Leiden algorithm, which we use to optimise modularity.
The retention strategy is never able to detect the planted partition correctly.
This is mostly because it finds more fine-grained structure within each planted cluster.
If we increase the average degree $\avg{k}$ (\figref{nmi_SBM_soc} middle), the retention strategy is also able to find the planted partition. However, the retention strategy requires a far larger degree $\avg{k}>15$ to do so than LPA, FLPA and the Leiden algorithm.

%For this we used a graph of $10^5$ nodes consisting of $100$ clusters and an average degree of $10$ with varying mixing parameter $\mu$.

\begin{figure*}
	\centering
    \includegraphics[width=\textwidth]{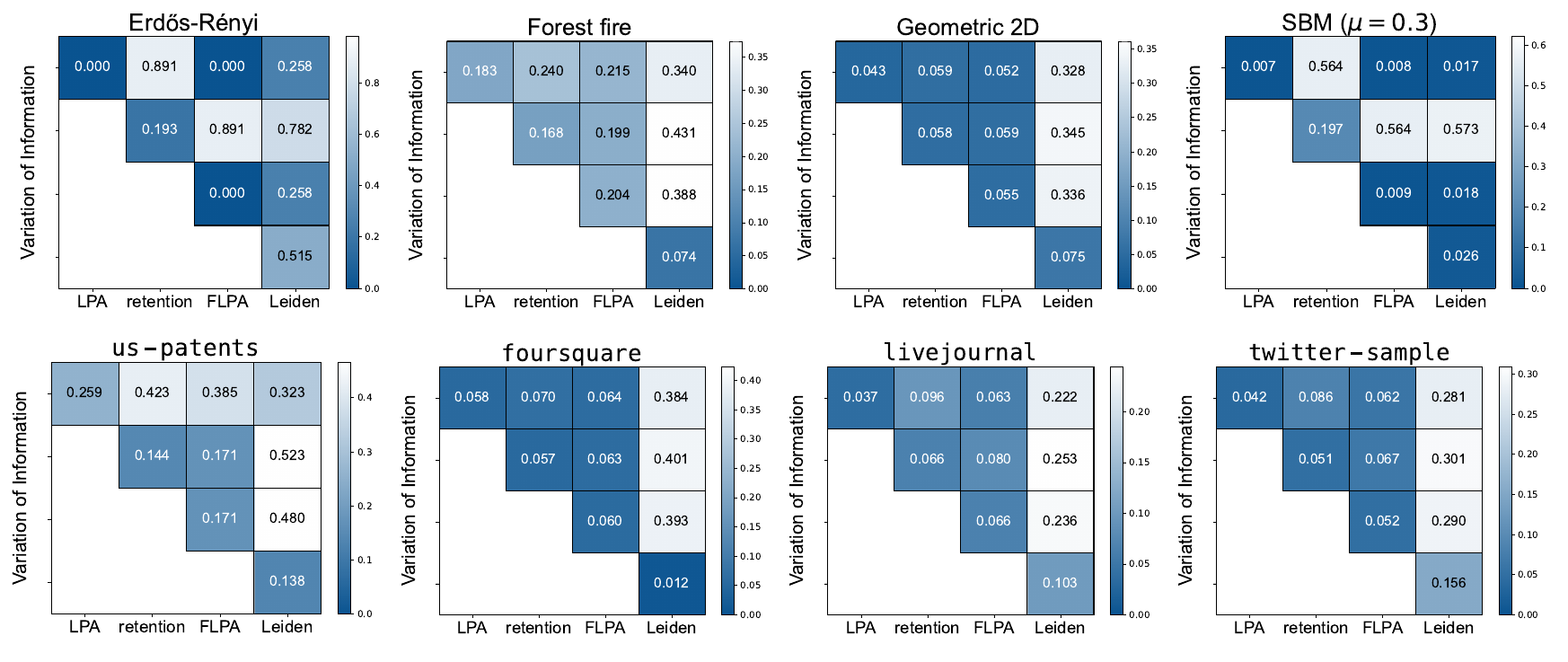}
	\caption{Distance between partitions of (top) synthetic networks with $n=10^5$ nodes and (bottom) empirical networks. % over $2\,000$ runs of the algorithms.
 }
	\label{fig:distance}
	% Check figures for some of the other figures for the real networks.
\end{figure*}

We also compare the differences across partitions that are detected by the different algorithms (Fig.~\ref{fig:distance} top) based on the normalised variation of information (VI)\cite{Meila2007-yk}.
The VI of an algorithm with itself denotes the average VI of two runs of the same algorithm on the exact same graph.
We can interpret this self-VI as a measure of stability: larger VI values suggest that partitions differ quite a bit from run to run, indicating a lower stability, while lower VI values suggest that partitions are quite similar from run to run, indicating a higher stability.
For ER graphs, there is essentially no variation within and between LPA and FLPA, because they always find a single large cluster.
The retention strategy, as explained earlier, typically finds some structure within ER graphs, which can also differ quite a bit from run to run.
The Leiden algorithm also shows quite different results from run to run, and also finds some structure within ER graphs, which is a known result for modularity\cite{Guimera2004-lv}.
For both the forest fire graphs and the two-dimensional geometric graphs, the three variants of LPA find relatively comparable structures, which differs from the partitions from the Leiden algorithm.
The stability of the partitions in the two-dimensional geometric graph of the three LPA variants is similar to the stability of the Leiden algorithm.
Finally, on SBM graphs, the retention strategy seems to be unable to find any meaningful partition.
As explained earlier, each group in the SBM is similar to an ER graph, where the retention strategy also finds very small substructures, which hence also shows up in the SBM results.
In contrast, both LPA and FLPA are able to find relatively stable partitions that are similar to each other, and similar to the results of the Leiden algorithm, as they all detect the planted partition of the SBM.

\subsubsection*{Empirical networks}

We now discuss the results of the empirical networks.
We first tested the algorithms on seven large empirical networks.
They vary in size, ranging from $317\,080$ nodes and $1\,049\,866$ edges for the smallest network (\texttt{com-dblp}), up to $6\,297\,539$ nodes and $16\,057\,711$ edges (\texttt{bitcoin}) or $4\,847\,571$ nodes and $68\,993\,773$ edges (\texttt{livejournal}).
FLPA is between 30--700 times faster than LPA (Table~\ref{tbl:time}) and between 4--15 times faster than the retention strategy.
For the largest network (\texttt{bitcoin}), LPA takes over $\SI{45}{\minute}$ on average, while FLPA is finished in $\SI{37}{\second}$.
LPA is by far the slowest on the \texttt{us-patents} network, where it takes over $\SI{7}{\hour}$, while FLPA is finished in $\SI{38}{\second}$.
This may be partly due to LPA finding a much coarser partition: the largest cluster covers more than $38\%$ of the nodes, and it finds about over $6$ times fewer clusters than the retention alternative, and over $3$ times fewer clusters than FLPA.
This conforms to the general pattern that LPA finds the least clusters, FLPA finds more clusters, while the retention strategy finds even more clusters.
There is no a priori reason to prefer larger over smaller clusters, so we cannot say whether a method would be preferable over the other based on this observation.

The partitions of most empirical networks themselves are similar between the three variants of LPA (Fig.~\ref{fig:distance} bottom).
Moreover, they differ similarly from the results of the Leiden algorithm.
The Leiden algorithm shows greater VI than (F)LPA for the \texttt{livejournal} and \texttt{twitter-sample} networks.
This might be related to the community sizes that Leiden finds, since we use it to optimise modularity, which suffers from a resolution limit\cite{Fortunato2007-bv}.
The resolution limit might lead the method to aggregate several clusters together, which may be somewhat arbitrary.
Indeed, the Leiden algorithm using modularity typically finds an order of magnitude fewer clusters than LPA.
As already noted earlier, for the \texttt{us-patents} network there is a larger difference between LPA and the retention variant and FLPA.

We also tested the algorithms on five small social networks with a known sociological division of nodes into communities (\figref{nmi_SBM_soc} right).
We compare the detected partitions with the node clusters described in the Methods.
We do not present the timing results for these networks, since they are too small for the measurements to be informative.
Overall, the results indicate that LPA finds the partitions with the highest NMI for these networks, followed by FLPA and then the retention strategy.
The Leiden algorithm, which optimises modularity, finds the worst partitions as measured by the NMI.
However, the results do not differ much between the algorithms.

\begin{table*}
	\centering
	\caption{\label{tbl:time}Speedup of fast label propagation for empirical networks. Results are averages over $2\,000$ runs of the algorithms ($1\,986$ runs for the \texttt{bitcoin} network).}
	\begin{tabular}{|lccccccc|}\hline
		Network & Nodes & Edges & Algorithm & Clusters & Largest & Time (s) & Speedup \\\hline
\multirow{3}{*}{\texttt{com-dblp}} & \multirow{3}{*}{$317\,080$} & \multirow{3}{*}{$1\,049\,866$} & LPA & $22\,048$ & $20.1\%$ & $185.0\pm105.0$ & $188.5\times$ \\
 &  &  & retention & $46\,380$ & $2.1\%$ & $10.6\pm6.0$ & $10.8\times$ \\
 &  &  & FLPA & $32\,329$ & $2.7\%$ & $1.0\pm0.4$ &  \\\hline
\multirow{3}{*}{\texttt{roadnet-ca}} & \multirow{3}{*}{$1\,965\,206$} & \multirow{3}{*}{$5\,533\,214$} & LPA & $219\,392$ & $0.0\%$ & $940.4\pm341.8$ & $161.8\times$ \\
 &  &  & retention & $610\,087$ & $0.0\%$ & $23.2\pm8.6$ & $4.0\times$ \\
 &  &  & FLPA & $341\,773$ & $0.0\%$ & $5.8\pm1.9$ &  \\\hline
\multirow{3}{*}{\texttt{us-patents}} & \multirow{3}{*}{$3\,774\,768$} & \multirow{3}{*}{$16\,522\,438$} & LPA & $53\,337$ & $38.3\%$ & $26\,704.4\pm12\,100.4$ & $705.2\times$ \\
 &  &  & retention & $359\,233$ & $1.4\%$ & $601.8\pm273.4$ & $15.9\times$ \\
 &  &  & FLPA & $183\,476$ & $1.8\%$ & $37.9\pm12.3$ &  \\\hline
\multirow{3}{*}{\texttt{foursquare}} & \multirow{3}{*}{$3\,935\,215$} & \multirow{3}{*}{$22\,809\,624$} & LPA & $73\,805$ & $6.2\%$ & $977.3\pm357.3$ & $63.8\times$ \\
 &  &  & retention & $77\,352$ & $5.3\%$ & $117.6\pm41.3$ & $7.7\times$ \\
 &  &  & FLPA & $76\,028$ & $5.8\%$ & $15.3\pm4.7$ &  \\\hline
\multirow{3}{*}{\texttt{livejournal}} & \multirow{3}{*}{$4\,847\,571$} & \multirow{3}{*}{$68\,993\,773$} & LPA & $59\,742$ & $69.4\%$ & $2\,248.1\pm1\,259.9$ & $30.2\times$ \\
 &  &  & retention & $149\,555$ & $60.5\%$ & $959.6\pm477.0$ & $12.9\times$ \\
 &  &  & FLPA & $93\,501$ & $65.5\%$ & $74.4\pm26.3$ &  \\\hline
\multirow{3}{*}{\texttt{twitter-sample}} & \multirow{3}{*}{$5\,384\,162$} & \multirow{3}{*}{$16\,011\,444$} & LPA & $16\,011$ & $55.1\%$ & $1\,343.5\pm544.6$ & $93.0\times$ \\
 &  &  & retention & $20\,651$ & $47.5\%$ & $92.8\pm38.9$ & $6.4\times$ \\
 &  &  & FLPA & $18\,560$ & $51.8\%$ & $14.5\pm4.9$ &  \\\hline
\multirow{3}{*}{\texttt{bitcoin}} & \multirow{3}{*}{$6\,297\,539$} & \multirow{3}{*}{$16\,057\,711$} & LPA & $42\,388$ & $50.9\%$ & $2\,937.7\pm1\,077.3$ & $80.3\times$ \\
 &  &  & retention & $1\,059\,801$ & $10.7\%$ & $597.1\pm522.6$ & $16.3\times$ \\
 &  &  & FLPA & $247\,404$ & $24.3\%$ & $36.6\pm13.0$ &  \\\hline
	\end{tabular}
\end{table*}

% \begin{figure}
% 	\centering
% 	\includegraphics[width=0.28\textwidth]{nmi-social}
% 	\caption{\label{fig:social}\redo{NMI of a few social smaller networks which have metadata on non-overlapping clusters.}}
% \end{figure}

\section*{Discussion}

\noindent
Detecting communities in networks is a frequent task in network analysis.
Label propagation is one of the fastest algorithms available.
It may be useful to get an initial first look at a network.
We here suggested a faster variant of label propagation.
Fast label propagation can run up to 700 times faster than the original label propagation.
Additionally, the same guarantees continue to hold, and results from both algorithms are largely comparable.
The quality of the partitions that FLPA finds seem to be on par with LPA.
When using label propagation, we believe our fast variant will bring benefits at no additional costs.
%It seems to be the fastest known algorithm for community detection.
Although label propagation may be useful to get a first look at a network, other methods may be likely to provide more accurate results.
One possibility is to use FLPA to obtain an initial rough partition, which is then further improved by the Leiden algorithm, aiming for a specific objective function.
Label propagation was also used to effectively and efficiently compress the Facebook graph\cite{BRSV11} to calculate its four degrees of separation\cite{BBRUV12}. 
At the same time, it is quite similar to majority opinion simulations~\cite{Lambiotte2007-vi}. 
The suggested speedup might also be relevant in the context of such applications.

\section*{Methods}

\subsection*{Algorithm implementation}

\noindent
We have implemented FLPA in C in \texttt{igraph} and made it available in its Python interface \texttt{python-igraph}.
The C source code can be found in \url{https://github.com/vtraag/python-igraph/tree/flpa}, while the Python interface can be found in \url{https://github.com/vtraag/igraph/tree/flpa}.
We compared FLPA to the existing implementation of LPA in (\texttt{python-})\texttt{igraph}.

\begin{algorithm}[ht]
\begin{algorithmic}[1]
	\REQUIRE undirected multigraph $G(V,E)$
	\ENSURE node clustering $C=\{c_i\}_{i=1}^n$
	\STATE $C\gets\{1,2,3\dots n\}$
	\WHILE{{\bf $C$ not maximal}}
		\FOR{$i\in\shuffle{V}$}
			\STATE $\{c\}\gets\argsmax_c\sum_{j\in V}A_{ij}\delta(c_j,c)$
			\STATE $c_i\gets\random_c\{c\}$
		\ENDFOR
	\ENDWHILE
	\RETURN $C$
\end{algorithmic}
\caption{Label propagation algorithm (LPA).}
\label{alg:lpa}
\end{algorithm}

\begin{algorithm}[ht]
\begin{algorithmic}[1]
	\REQUIRE undirected multigraph $G(V,E)$
	\ENSURE node clustering $C=\{c_i\}_{i=1}^n$
	\STATE $C\gets\{1,2,3\dots n\}$
	\WHILE{{\bf $C$ changed}}
		\FOR{$i\in\shuffle{V}$}
			\STATE $\{c\}\gets\argsmax_c\sum_{j\in V}A_{ij}\delta(c_j,c)$
			\IF{$c_i\notin \{c\}$}
				\STATE $c_i\gets\random_c\{c\}$
			\ENDIF
		\ENDFOR
	\ENDWHILE
	\RETURN $C$
\end{algorithmic}
\caption{Label propagation algorithm with retention.}
\label{alg:lpar}
\end{algorithm}

\begin{algorithm}[ht]
\begin{algorithmic}[1]
	\REQUIRE undirected multigraph $G(V,E)$
	\ENSURE node clustering $C=\{c_i\}_{i=1}^n$
	\STATE $C\gets\{1,2,3\dots n\}$
	\STATE $Q\gets\shuffle{V}$
	\WHILE{{\bf $Q$ not empty}}
		\STATE $i\gets\pops{Q}$
		\STATE $\{c\}\gets\argsmax_c\sum_{j\in V}A_{ij}\delta(c_j,c)$
		\STATE $c\gets\random_c\{c\}$
		\IF{$c_i\neq c$}
			\STATE $c_i\gets c$
			\FOR{$j\in N_i: c_j\neq c\wedge j\notin Q$}
				\STATE $j\rightarrow\pushs{Q}$
			\ENDFOR
		\ENDIF
	\ENDWHILE
	\RETURN $C$
\end{algorithmic}
\caption{Fast label propagation algorithm (FLPA).}
\label{alg:flpa}
\end{algorithm}

\subsection*{Empirical networks}

The large empirical networks from Table~\ref{tbl:time} are part of the Netzschleuder repository and can be downloaded from \url{https://networks.skewed.de}. 
All networks have been reduced to their largest connected component.
The \texttt{com-dblp} is a co-authorship network extracted from the DBLP database in 2012\cite{YL12}, 
the \texttt{roadnet-ca} is the road network of California\cite{leskovec_community_2009}, 
the \texttt{us-patents} is the U.S. patent citation network from 1975 to 1999\cite{hall_nber_2001},
the \texttt{foursquare} network represents check-in events on Foursquare from April 2012 to September 2013\cite{YANG2015170},
the \texttt{livejournal}
is an online social network of the LiveJournal members in 2006\cite{BHKL06},
the \texttt{twitter-sample} is a sample of the Twitter follower network extracted in 2012\cite{twitter},
and the \texttt{bitcoin}
is a network of Bitcoin transactions from January 2009 to April 2013\cite{bitcoin}.

The small social networks from Figure~\ref{fig:nmi_SBM_soc} are either part of the Netzschleuder repository or available as supplementary material from Ref.\cite{primary_school}.
The \texttt{karate} is a friendship network among members of a university karate club divided into two factions\cite{karate_club} ($34$ nodes, $78$ edges).
The \texttt{dolphins} is a social network of frequent associations observed among dolphins living off New Zealand\cite{lusseau_bottlenose}, with a sociological division of dolphins into two groups ($62$ nodes, $159$ edges).
The \texttt{football} network represents American football games between U.S.\ colleges during the 2000 regular season\cite{girvan_community_2002}, with each college assigned to one of twelve conferences ($115$ nodes, $616$ edges).
The \texttt{school-day1} and \texttt{school-day2} networks encode face-to-face interactions between children and teachers in a French elementary school on two consecutive days\cite{primary_school}, where the metadata contain the assignment of children to 10 classes. Following the original study, we only include edges between individuals who interacted for at least 2 minutes ($236$ and $238$ nodes, $1\,956$ and $2\,177$ edges, respectively).

\section*{Data availability}

The code created for the current study is implemented in the \texttt{igraph} library, available from \url{https://github.com/vtraag/igraph/tree/flpa}.
The datasets analysed during the current study are available in the Netzschleuder repository, \url{https://networks.skewed.de}, or via references in the published article.

\bibliography{bibliography.bib}

%aipnum4-2.bst 2019-01-14 (MD) hand-edited version of apsrev4-1.bst
\begin{thebibliography}{40}%
\makeatletter
\providecommand \@ifxundefined [1]{%
 \@ifx{#1\undefined}
}%
\providecommand \@ifnum [1]{%
 \ifnum #1\expandafter \@firstoftwo
 \else \expandafter \@secondoftwo
 \fi
}%
\providecommand \@ifx [1]{%
 \ifx #1\expandafter \@firstoftwo
 \else \expandafter \@secondoftwo
 \fi
}%
\providecommand \natexlab [1]{#1}%
\providecommand \enquote  [1]{``#1''}%
\providecommand \bibnamefont  [1]{#1}%
\providecommand \bibfnamefont [1]{#1}%
\providecommand \citenamefont [1]{#1}%
\providecommand \href@noop [0]{\@secondoftwo}%
\providecommand \href [0]{\begingroup \@sanitize@url \@href}%
\providecommand \@href[1]{\@@startlink{#1}\@@href}%
\providecommand \@@href[1]{\endgroup#1\@@endlink}%
\providecommand \@sanitize@url [0]{\catcode `\\12\catcode `\$12\catcode
  `\&12\catcode `\#12\catcode `\^12\catcode `\_12\catcode `\%12\relax}%
\providecommand \@@startlink[1]{}%
\providecommand \@@endlink[0]{}%
\providecommand \url  [0]{\begingroup\@sanitize@url \@url }%
\providecommand \@url [1]{\endgroup\@href {#1}{\urlprefix }}%
\providecommand \urlprefix  [0]{URL }%
\providecommand \Eprint [0]{\href }%
\providecommand \doibase [0]{https://doi.org/}%
\providecommand \selectlanguage [0]{\@gobble}%
\providecommand \bibinfo  [0]{\@secondoftwo}%
\providecommand \bibfield  [0]{\@secondoftwo}%
\providecommand \translation [1]{[#1]}%
\providecommand \BibitemOpen [0]{}%
\providecommand \bibitemStop [0]{}%
\providecommand \bibitemNoStop [0]{.\EOS\space}%
\providecommand \EOS [0]{\spacefactor3000\relax}%
\providecommand \BibitemShut  [1]{\csname bibitem#1\endcsname}%
\let\auto@bib@innerbib\@empty
%</preamble>
\bibitem [{\citenamefont {Newman}\ and\ \citenamefont {Girvan}(2004)}]{NG04}%
  \BibitemOpen
  \bibfield  {author} {\bibinfo {author} {\bibfnamefont {M.~E.~J.}\
  \bibnamefont {Newman}}\ and\ \bibinfo {author} {\bibfnamefont
  {M.}~\bibnamefont {Girvan}},\ }\bibfield  {title} {\enquote {\bibinfo {title}
  {Finding and evaluating community structure in networks},}\ }\href@noop {}
  {\bibfield  {journal} {\bibinfo  {journal} {Phys. Rev. E}\ }\textbf {\bibinfo
  {volume} {69}},\ \bibinfo {pages} {026113} (\bibinfo {year}
  {2004})}\BibitemShut {NoStop}%
\bibitem [{\citenamefont {Peixoto}(2020)}]{Pei20}%
  \BibitemOpen
  \bibfield  {author} {\bibinfo {author} {\bibfnamefont {T.~P.}\ \bibnamefont
  {Peixoto}},\ }\bibfield  {title} {\enquote {\bibinfo {title} {Bayesian
  stochastic blockmodeling},}\ }in\ \href {http://arxiv.org/abs/1705.10225}
  {\emph {\bibinfo {booktitle} {Advances in {Network} {Clustering} and
  {Blockmodeling}}}},\ \bibinfo {series and number} {Computational and
  {Quantitative} {Social} {Science}},\ \bibinfo {editor} {edited by\ \bibinfo
  {editor} {\bibfnamefont {P.}~\bibnamefont {Doreian}}, \bibinfo {editor}
  {\bibfnamefont {V.}~\bibnamefont {Batagelj}},\ and\ \bibinfo {editor}
  {\bibfnamefont {A.}~\bibnamefont {Ferligoj}}}\ (\bibinfo  {publisher}
  {Wiley},\ \bibinfo {address} {New York},\ \bibinfo {year} {2020})\ \bibinfo
  {edition} {1st}\ ed.,\ pp.\ \bibinfo {pages} {281--324}\BibitemShut {NoStop}%
\bibitem [{\citenamefont {Rosvall}\ and\ \citenamefont
  {Bergstrom}(2007)}]{RB07}%
  \BibitemOpen
  \bibfield  {author} {\bibinfo {author} {\bibfnamefont {M.}~\bibnamefont
  {Rosvall}}\ and\ \bibinfo {author} {\bibfnamefont {C.~T.}\ \bibnamefont
  {Bergstrom}},\ }\bibfield  {title} {\enquote {\bibinfo {title} {An
  information-theoretic framework for resolving community structure in complex
  networks},}\ }\href@noop {} {\bibfield  {journal} {\bibinfo  {journal} {P.
  Natl. Acad. Sci. USA}\ }\textbf {\bibinfo {volume} {104}},\ \bibinfo {pages}
  {7327--7331} (\bibinfo {year} {2007})}\BibitemShut {NoStop}%
\bibitem [{\citenamefont {Rosvall}\ and\ \citenamefont
  {Bergstrom}(2008)}]{RB08}%
  \BibitemOpen
  \bibfield  {author} {\bibinfo {author} {\bibfnamefont {M.}~\bibnamefont
  {Rosvall}}\ and\ \bibinfo {author} {\bibfnamefont {C.~T.}\ \bibnamefont
  {Bergstrom}},\ }\bibfield  {title} {\enquote {\bibinfo {title} {Maps of
  random walks on complex networks reveal community structure},}\ }\href@noop
  {} {\bibfield  {journal} {\bibinfo  {journal} {Proc. Natl. Acad. Sci. U. S.
  A.}\ }\textbf {\bibinfo {volume} {105}},\ \bibinfo {pages} {1118--1123}
  (\bibinfo {year} {2008})}\BibitemShut {NoStop}%
\bibitem [{\citenamefont {Clauset}, \citenamefont {Newman},\ and\ \citenamefont
  {Moore}(2004)}]{Clauset2004-bf}%
  \BibitemOpen
  \bibfield  {author} {\bibinfo {author} {\bibfnamefont {A.}~\bibnamefont
  {Clauset}}, \bibinfo {author} {\bibfnamefont {M.~E.~J.}\ \bibnamefont
  {Newman}},\ and\ \bibinfo {author} {\bibfnamefont {C.}~\bibnamefont
  {Moore}},\ }\bibfield  {title} {\enquote {\bibinfo {title} {Finding community
  structure in very large networks},}\ }\href@noop {} {\bibfield  {journal}
  {\bibinfo  {journal} {Physical Review E}\ }\textbf {\bibinfo {volume} {70}},\
  \bibinfo {pages} {066111} (\bibinfo {year} {2004})}\BibitemShut {NoStop}%
\bibitem [{\citenamefont {Reichardt}\ and\ \citenamefont
  {Bornholdt}(2006)}]{Reichardt2006-ij}%
  \BibitemOpen
  \bibfield  {author} {\bibinfo {author} {\bibfnamefont {J.}~\bibnamefont
  {Reichardt}}\ and\ \bibinfo {author} {\bibfnamefont {S.}~\bibnamefont
  {Bornholdt}},\ }\bibfield  {title} {\enquote {\bibinfo {title} {Statistical
  mechanics of community detection},}\ }\href@noop {} {\bibfield  {journal}
  {\bibinfo  {journal} {Physical Review E}\ }\textbf {\bibinfo {volume} {74}},\
  \bibinfo {pages} {016110} (\bibinfo {year} {2006})}\BibitemShut {NoStop}%
\bibitem [{\citenamefont {Duch}\ and\ \citenamefont
  {Arenas}(2005)}]{Duch2005-jj}%
  \BibitemOpen
  \bibfield  {author} {\bibinfo {author} {\bibfnamefont {J.}~\bibnamefont
  {Duch}}\ and\ \bibinfo {author} {\bibfnamefont {A.}~\bibnamefont {Arenas}},\
  }\bibfield  {title} {\enquote {\bibinfo {title} {Community detection in
  complex networks using extremal optimization},}\ }\href@noop {} {\bibfield
  {journal} {\bibinfo  {journal} {Physical Review E}\ }\textbf {\bibinfo
  {volume} {72}},\ \bibinfo {pages} {027104} (\bibinfo {year}
  {2005})}\BibitemShut {NoStop}%
\bibitem [{\citenamefont {Blondel}\ \emph {et~al.}(2008)\citenamefont
  {Blondel}, \citenamefont {Guillaume}, \citenamefont {Lambiotte},\ and\
  \citenamefont {Lefebvre}}]{BGLL08}%
  \BibitemOpen
  \bibfield  {author} {\bibinfo {author} {\bibfnamefont {V.~D.}\ \bibnamefont
  {Blondel}}, \bibinfo {author} {\bibfnamefont {J.-L.}\ \bibnamefont
  {Guillaume}}, \bibinfo {author} {\bibfnamefont {R.}~\bibnamefont
  {Lambiotte}},\ and\ \bibinfo {author} {\bibfnamefont {E.}~\bibnamefont
  {Lefebvre}},\ }\bibfield  {title} {\enquote {\bibinfo {title} {Fast unfolding
  of communities in large networks},}\ }\href@noop {} {\bibfield  {journal}
  {\bibinfo  {journal} {J. Stat. Mech.}\ }\textbf {\bibinfo {volume} {P10008}}
  (\bibinfo {year} {2008})}\BibitemShut {NoStop}%
\bibitem [{\citenamefont {Traag}, \citenamefont {Waltman},\ and\ \citenamefont
  {Van~Eck}(2019)}]{TWV19}%
  \BibitemOpen
  \bibfield  {author} {\bibinfo {author} {\bibfnamefont {V.~A.}\ \bibnamefont
  {Traag}}, \bibinfo {author} {\bibfnamefont {L.}~\bibnamefont {Waltman}},\
  and\ \bibinfo {author} {\bibfnamefont {N.~J.}\ \bibnamefont {Van~Eck}},\
  }\bibfield  {title} {\enquote {\bibinfo {title} {From {Louvain} to {Leiden}:
  {Guaranteeing} well-connected communities},}\ }\href@noop {} {\bibfield
  {journal} {\bibinfo  {journal} {Sci. Rep.}\ }\textbf {\bibinfo {volume}
  {9}},\ \bibinfo {pages} {5233} (\bibinfo {year} {2019})}\BibitemShut
  {NoStop}%
\bibitem [{\citenamefont {Traag}, \citenamefont {Van~Dooren},\ and\
  \citenamefont {Nesterov}(2011)}]{Traag2011-nu}%
  \BibitemOpen
  \bibfield  {author} {\bibinfo {author} {\bibfnamefont {V.~A.}\ \bibnamefont
  {Traag}}, \bibinfo {author} {\bibfnamefont {P.}~\bibnamefont {Van~Dooren}},\
  and\ \bibinfo {author} {\bibfnamefont {Y.}~\bibnamefont {Nesterov}},\
  }\bibfield  {title} {\enquote {\bibinfo {title} {Narrow scope for
  resolution-limit-free community detection},}\ }\href@noop {} {\bibfield
  {journal} {\bibinfo  {journal} {Physical Review E}\ }\textbf {\bibinfo
  {volume} {84}},\ \bibinfo {pages} {016114} (\bibinfo {year}
  {2011})}\BibitemShut {NoStop}%
\bibitem [{\citenamefont {Raghavan}, \citenamefont {Albert},\ and\
  \citenamefont {Kumara}(2007)}]{RAK07}%
  \BibitemOpen
  \bibfield  {author} {\bibinfo {author} {\bibfnamefont {U.~N.}\ \bibnamefont
  {Raghavan}}, \bibinfo {author} {\bibfnamefont {R.}~\bibnamefont {Albert}},\
  and\ \bibinfo {author} {\bibfnamefont {S.}~\bibnamefont {Kumara}},\
  }\bibfield  {title} {\enquote {\bibinfo {title} {Near linear time algorithm
  to detect community structures in large-scale networks},}\ }\href@noop {}
  {\bibfield  {journal} {\bibinfo  {journal} {Phys. Rev. E}\ }\textbf {\bibinfo
  {volume} {76}},\ \bibinfo {pages} {036106} (\bibinfo {year}
  {2007})}\BibitemShut {NoStop}%
\bibitem [{\citenamefont {{\v S}ubelj}(2020)}]{Sub20}%
  \BibitemOpen
  \bibfield  {author} {\bibinfo {author} {\bibfnamefont {L.}~\bibnamefont {{\v
  S}ubelj}},\ }\bibfield  {title} {\enquote {\bibinfo {title} {Label
  propagation for clustering},}\ }in\ \href@noop {} {\emph {\bibinfo
  {booktitle} {Advances in {Network} {Clustering} and {Blockmodeling}}}},\
  \bibinfo {series and number} {Computational and {Quantitative} {Social}
  {Science}},\ \bibinfo {editor} {edited by\ \bibinfo {editor} {\bibfnamefont
  {P.}~\bibnamefont {Doreian}}, \bibinfo {editor} {\bibfnamefont
  {V.}~\bibnamefont {Batagelj}},\ and\ \bibinfo {editor} {\bibfnamefont
  {A.}~\bibnamefont {Ferligoj}}}\ (\bibinfo  {publisher} {Wiley},\ \bibinfo
  {address} {New York},\ \bibinfo {year} {2020})\ \bibinfo {edition} {1st}\
  ed.,\ pp.\ \bibinfo {pages} {121--150}\BibitemShut {NoStop}%
\bibitem [{\citenamefont {Tib{\'e}ly}\ and\ \citenamefont
  {Kert{\'e}sz}(2008)}]{Tibely2008-qd}%
  \BibitemOpen
  \bibfield  {author} {\bibinfo {author} {\bibfnamefont {G.}~\bibnamefont
  {Tib{\'e}ly}}\ and\ \bibinfo {author} {\bibfnamefont {J.}~\bibnamefont
  {Kert{\'e}sz}},\ }\bibfield  {title} {\enquote {\bibinfo {title} {On the
  equivalence of the label propagation method of community detection and a
  potts model approach},}\ }\href@noop {} {\bibfield  {journal} {\bibinfo
  {journal} {Physica A: Statistical Mechanics and its Applications}\ }\textbf
  {\bibinfo {volume} {387}},\ \bibinfo {pages} {4982--4984} (\bibinfo {year}
  {2008})}\BibitemShut {NoStop}%
\bibitem [{\citenamefont {Garza}\ and\ \citenamefont
  {Schaeffer}(2019)}]{Garza2019-td}%
  \BibitemOpen
  \bibfield  {author} {\bibinfo {author} {\bibfnamefont {S.~E.}\ \bibnamefont
  {Garza}}\ and\ \bibinfo {author} {\bibfnamefont {S.~E.}\ \bibnamefont
  {Schaeffer}},\ }\bibfield  {title} {\enquote {\bibinfo {title} {Community
  detection with the label propagation algorithm: A survey},}\ }\href@noop {}
  {\bibfield  {journal} {\bibinfo  {journal} {Physica A: Statistical Mechanics
  and its Applications}\ }\textbf {\bibinfo {volume} {534}},\ \bibinfo {pages}
  {122058} (\bibinfo {year} {2019})}\BibitemShut {NoStop}%
\bibitem [{\citenamefont {Radicchi}\ \emph {et~al.}(2004)\citenamefont
  {Radicchi}, \citenamefont {Castellano}, \citenamefont {Cecconi},
  \citenamefont {Loreto},\ and\ \citenamefont {Parisi}}]{Radicchi2004-yn}%
  \BibitemOpen
  \bibfield  {author} {\bibinfo {author} {\bibfnamefont {F.}~\bibnamefont
  {Radicchi}}, \bibinfo {author} {\bibfnamefont {C.}~\bibnamefont
  {Castellano}}, \bibinfo {author} {\bibfnamefont {F.}~\bibnamefont {Cecconi}},
  \bibinfo {author} {\bibfnamefont {V.}~\bibnamefont {Loreto}},\ and\ \bibinfo
  {author} {\bibfnamefont {D.}~\bibnamefont {Parisi}},\ }\bibfield  {title}
  {\enquote {\bibinfo {title} {Defining and identifying communities in
  networks},}\ }\href@noop {} {\bibfield  {journal} {\bibinfo  {journal}
  {Proceedings of the National Academy of Sciences}\ }\textbf {\bibinfo
  {volume} {101}},\ \bibinfo {pages} {2658--2663} (\bibinfo {year}
  {2004})}\BibitemShut {NoStop}%
\bibitem [{\citenamefont {Barber}\ and\ \citenamefont
  {Clark}(2009)}]{Barber2009-vb}%
  \BibitemOpen
  \bibfield  {author} {\bibinfo {author} {\bibfnamefont {M.~J.}\ \bibnamefont
  {Barber}}\ and\ \bibinfo {author} {\bibfnamefont {J.~W.}\ \bibnamefont
  {Clark}},\ }\bibfield  {title} {\enquote {\bibinfo {title} {Detecting network
  communities by propagating labels under constraints},}\ }\href@noop {}
  {\bibfield  {journal} {\bibinfo  {journal} {Phys. Rev. E}\ }\textbf {\bibinfo
  {volume} {80}},\ \bibinfo {pages} {026129} (\bibinfo {year}
  {2009})}\BibitemShut {NoStop}%
\bibitem [{\citenamefont {Leung}\ \emph {et~al.}(2009)\citenamefont {Leung},
  \citenamefont {Hui}, \citenamefont {Li{\`o}},\ and\ \citenamefont
  {Crowcroft}}]{LHLC09}%
  \BibitemOpen
  \bibfield  {author} {\bibinfo {author} {\bibfnamefont {I.~X.~Y.}\
  \bibnamefont {Leung}}, \bibinfo {author} {\bibfnamefont {P.}~\bibnamefont
  {Hui}}, \bibinfo {author} {\bibfnamefont {P.}~\bibnamefont {Li{\`o}}},\ and\
  \bibinfo {author} {\bibfnamefont {J.}~\bibnamefont {Crowcroft}},\ }\bibfield
  {title} {\enquote {\bibinfo {title} {Towards real-time community detection in
  large networks},}\ }\href@noop {} {\bibfield  {journal} {\bibinfo  {journal}
  {Phys. Rev. E}\ }\textbf {\bibinfo {volume} {79}},\ \bibinfo {pages} {066107}
  (\bibinfo {year} {2009})}\BibitemShut {NoStop}%
\bibitem [{\citenamefont {{\v S}ubelj}\ and\ \citenamefont
  {Bajec}(2011)}]{Subelj2011-cb}%
  \BibitemOpen
  \bibfield  {author} {\bibinfo {author} {\bibfnamefont {L.}~\bibnamefont {{\v
  S}ubelj}}\ and\ \bibinfo {author} {\bibfnamefont {M.}~\bibnamefont {Bajec}},\
  }\bibfield  {title} {\enquote {\bibinfo {title} {Unfolding communities in
  large complex networks: Combining defensive and offensive label propagation
  for core extraction},}\ }\href@noop {} {\bibfield  {journal} {\bibinfo
  {journal} {Phys. Rev. E}\ }\textbf {\bibinfo {volume} {83}},\ \bibinfo
  {pages} {036103} (\bibinfo {year} {2011})}\BibitemShut {NoStop}%
\bibitem [{\citenamefont {Xie}\ and\ \citenamefont {Szymanski}(2011)}]{XS11}%
  \BibitemOpen
  \bibfield  {author} {\bibinfo {author} {\bibfnamefont {J.}~\bibnamefont
  {Xie}}\ and\ \bibinfo {author} {\bibfnamefont {B.~K.}\ \bibnamefont
  {Szymanski}},\ }\bibfield  {title} {\enquote {\bibinfo {title} {Community
  detection using a neighborhood strength driven label propagation
  algorithm},}\ }in\ \href@noop {} {\emph {\bibinfo {booktitle} {Proceedings of
  the {IEEE} {International} {Workshop} on {Network} {Science}}}}\ (\bibinfo
  {address} {West Point, NY, USA},\ \bibinfo {year} {2011})\ pp.\ \bibinfo
  {pages} {188--195}\BibitemShut {NoStop}%
\bibitem [{\citenamefont {Tasgin}\ and\ \citenamefont
  {Bingol}(2019)}]{TASGIN2019315}%
  \BibitemOpen
  \bibfield  {author} {\bibinfo {author} {\bibfnamefont {M.}~\bibnamefont
  {Tasgin}}\ and\ \bibinfo {author} {\bibfnamefont {H.~O.}\ \bibnamefont
  {Bingol}},\ }\bibfield  {title} {\enquote {\bibinfo {title} {Community
  detection using boundary nodes in complex networks},}\ }\href@noop {}
  {\bibfield  {journal} {\bibinfo  {journal} {Physica A: Statistical Mechanics
  and its Applications}\ }\textbf {\bibinfo {volume} {513}},\ \bibinfo {pages}
  {315--324} (\bibinfo {year} {2019})}\BibitemShut {NoStop}%
\bibitem [{\citenamefont {Newman}(2018)}]{newman_networks_2018}%
  \BibitemOpen
  \bibfield  {author} {\bibinfo {author} {\bibfnamefont {M.~E.~J.}\
  \bibnamefont {Newman}},\ }\href@noop {} {\emph {\bibinfo {title}
  {Networks}}},\ \bibinfo {edition} {2nd}\ ed.\ (\bibinfo  {publisher} {Oxford
  University Press},\ \bibinfo {address} {Oxford},\ \bibinfo {year}
  {2018})\BibitemShut {NoStop}%
\bibitem [{\citenamefont {Danon}\ \emph {et~al.}(2005)\citenamefont {Danon},
  \citenamefont {D{\'i}az-Guilera}, \citenamefont {Duch},\ and\ \citenamefont
  {Arenas}}]{danon_comparing_2005}%
  \BibitemOpen
  \bibfield  {author} {\bibinfo {author} {\bibfnamefont {L.}~\bibnamefont
  {Danon}}, \bibinfo {author} {\bibfnamefont {A.}~\bibnamefont
  {D{\'i}az-Guilera}}, \bibinfo {author} {\bibfnamefont {J.}~\bibnamefont
  {Duch}},\ and\ \bibinfo {author} {\bibfnamefont {A.}~\bibnamefont {Arenas}},\
  }\bibfield  {title} {\enquote {\bibinfo {title} {Comparing community
  structure identification},}\ }\href@noop {} {\bibfield  {journal} {\bibinfo
  {journal} {J. Stat. Mech.}\ ,\ \bibinfo {pages} {P09008}} (\bibinfo {year}
  {2005})}\BibitemShut {NoStop}%
\bibitem [{\citenamefont {Floretta}\ \emph {et~al.}(2013)\citenamefont
  {Floretta}, \citenamefont {Liechti}, \citenamefont {Flammini},\ and\
  \citenamefont {De~Los~Rios}}]{Floretta2013-zh}%
  \BibitemOpen
  \bibfield  {author} {\bibinfo {author} {\bibfnamefont {L.}~\bibnamefont
  {Floretta}}, \bibinfo {author} {\bibfnamefont {J.}~\bibnamefont {Liechti}},
  \bibinfo {author} {\bibfnamefont {A.}~\bibnamefont {Flammini}},\ and\
  \bibinfo {author} {\bibfnamefont {P.}~\bibnamefont {De~Los~Rios}},\
  }\bibfield  {title} {\enquote {\bibinfo {title} {Stochastic fluctuations and
  the detectability limit of network communities},}\ }\href
  {https://doi.org/10.1103/PhysRevE.88.060801} {\bibfield  {journal} {\bibinfo
  {journal} {Phys. Rev. E}\ }\textbf {\bibinfo {volume} {88}},\ \bibinfo
  {pages} {060801} (\bibinfo {year} {2013})}\BibitemShut {NoStop}%
\bibitem [{\citenamefont {Meil{\u a}}(2007)}]{Meila2007-yk}%
  \BibitemOpen
  \bibfield  {author} {\bibinfo {author} {\bibfnamefont {M.}~\bibnamefont
  {Meil{\u a}}},\ }\bibfield  {title} {\enquote {\bibinfo {title} {Comparing
  clusterings: An information based distance},}\ }\href@noop {} {\bibfield
  {journal} {\bibinfo  {journal} {J. Multivar. Anal.}\ }\textbf {\bibinfo
  {volume} {98}},\ \bibinfo {pages} {873--895} (\bibinfo {year}
  {2007})}\BibitemShut {NoStop}%
\bibitem [{\citenamefont {Guimer{\`a}}, \citenamefont {Sales-Pardo},\ and\
  \citenamefont {Amaral}(2004)}]{Guimera2004-lv}%
  \BibitemOpen
  \bibfield  {author} {\bibinfo {author} {\bibfnamefont {R.}~\bibnamefont
  {Guimer{\`a}}}, \bibinfo {author} {\bibfnamefont {M.}~\bibnamefont
  {Sales-Pardo}},\ and\ \bibinfo {author} {\bibfnamefont {L.}~\bibnamefont
  {Amaral}},\ }\bibfield  {title} {\enquote {\bibinfo {title} {Modularity from
  fluctuations in random graphs and complex networks},}\ }\href@noop {}
  {\bibfield  {journal} {\bibinfo  {journal} {Physical Review E}\ }\textbf
  {\bibinfo {volume} {70}},\ \bibinfo {pages} {025101} (\bibinfo {year}
  {2004})}\BibitemShut {NoStop}%
\bibitem [{\citenamefont {Fortunato}\ and\ \citenamefont
  {Barth{\'e}lemy}(2007)}]{Fortunato2007-bv}%
  \BibitemOpen
  \bibfield  {author} {\bibinfo {author} {\bibfnamefont {S.}~\bibnamefont
  {Fortunato}}\ and\ \bibinfo {author} {\bibfnamefont {M.}~\bibnamefont
  {Barth{\'e}lemy}},\ }\bibfield  {title} {\enquote {\bibinfo {title}
  {Resolution limit in community detection},}\ }\href@noop {} {\bibfield
  {journal} {\bibinfo  {journal} {Proc. Natl. Acad. Sci.}\ }\textbf {\bibinfo
  {volume} {104}},\ \bibinfo {pages} {36} (\bibinfo {year} {2007})}\BibitemShut
  {NoStop}%
\bibitem [{\citenamefont {Boldi}\ \emph {et~al.}(2011)\citenamefont {Boldi},
  \citenamefont {Rosa}, \citenamefont {Santini},\ and\ \citenamefont
  {Vigna}}]{BRSV11}%
  \BibitemOpen
  \bibfield  {author} {\bibinfo {author} {\bibfnamefont {P.}~\bibnamefont
  {Boldi}}, \bibinfo {author} {\bibfnamefont {M.}~\bibnamefont {Rosa}},
  \bibinfo {author} {\bibfnamefont {M.}~\bibnamefont {Santini}},\ and\ \bibinfo
  {author} {\bibfnamefont {S.}~\bibnamefont {Vigna}},\ }\bibfield  {title}
  {\enquote {\bibinfo {title} {Layered label propagation: {A} multiresolution
  coordinate-free ordering for compressing social networks},}\ }in\ \href@noop
  {} {\emph {\bibinfo {booktitle} {Proceedings of the {International} {World}
  {Wide} {Web} {Conference}}}}\ (\bibinfo {address} {Hyderabad, India},\
  \bibinfo {year} {2011})\ pp.\ \bibinfo {pages} {587--596}\BibitemShut
  {NoStop}%
\bibitem [{\citenamefont {Backstrom}\ \emph {et~al.}(2012)\citenamefont
  {Backstrom}, \citenamefont {Boldi}, \citenamefont {Rosa}, \citenamefont
  {Ugander},\ and\ \citenamefont {Vigna}}]{BBRUV12}%
  \BibitemOpen
  \bibfield  {author} {\bibinfo {author} {\bibfnamefont {L.}~\bibnamefont
  {Backstrom}}, \bibinfo {author} {\bibfnamefont {P.}~\bibnamefont {Boldi}},
  \bibinfo {author} {\bibfnamefont {M.}~\bibnamefont {Rosa}}, \bibinfo {author}
  {\bibfnamefont {J.}~\bibnamefont {Ugander}},\ and\ \bibinfo {author}
  {\bibfnamefont {S.}~\bibnamefont {Vigna}},\ }\bibfield  {title} {\enquote
  {\bibinfo {title} {Four degrees of separation},}\ }in\ \href@noop {} {\emph
  {\bibinfo {booktitle} {Proceedings of the {ACM} {International} {Conference}
  on {Web} {Science}}}}\ (\bibinfo {address} {Evanston, IL, USA},\ \bibinfo
  {year} {2012})\ pp.\ \bibinfo {pages} {45--54}\BibitemShut {NoStop}%
\bibitem [{\citenamefont {Lambiotte}\ and\ \citenamefont
  {Ausloos}(2007)}]{Lambiotte2007-vi}%
  \BibitemOpen
  \bibfield  {author} {\bibinfo {author} {\bibfnamefont {R.}~\bibnamefont
  {Lambiotte}}\ and\ \bibinfo {author} {\bibfnamefont {M.}~\bibnamefont
  {Ausloos}},\ }\bibfield  {title} {\enquote {\bibinfo {title} {Coexistence of
  opposite opinions in a network with communities},}\ }\href@noop {} {\bibfield
   {journal} {\bibinfo  {journal} {Journal of Statistical Mechanics: Theory and
  Experiment}\ }\textbf {\bibinfo {volume} {2007}},\ \bibinfo {pages} {P08026}
  (\bibinfo {year} {2007})}\BibitemShut {NoStop}%
\bibitem [{\citenamefont {Yang}\ and\ \citenamefont {Leskovec}(2012)}]{YL12}%
  \BibitemOpen
  \bibfield  {author} {\bibinfo {author} {\bibfnamefont {J.}~\bibnamefont
  {Yang}}\ and\ \bibinfo {author} {\bibfnamefont {J.}~\bibnamefont
  {Leskovec}},\ }\bibfield  {title} {\enquote {\bibinfo {title} {Defining and
  evaluating network communities based on ground-truth},}\ }in\ \href@noop {}
  {\emph {\bibinfo {booktitle} {Proceedings of the ACM SIGKDD Workshop on
  Mining Data Semantics}}}\ (\bibinfo {address} {Beijing, China},\ \bibinfo
  {year} {2012})\ pp.\ \bibinfo {pages} {1--8}\BibitemShut {NoStop}%
\bibitem [{\citenamefont {Leskovec}\ \emph {et~al.}(2009)\citenamefont
  {Leskovec}, \citenamefont {Lang}, \citenamefont {Dasgupta},\ and\
  \citenamefont {Mahoney}}]{leskovec_community_2009}%
  \BibitemOpen
  \bibfield  {author} {\bibinfo {author} {\bibfnamefont {J.}~\bibnamefont
  {Leskovec}}, \bibinfo {author} {\bibfnamefont {K.~J.}\ \bibnamefont {Lang}},
  \bibinfo {author} {\bibfnamefont {A.}~\bibnamefont {Dasgupta}},\ and\
  \bibinfo {author} {\bibfnamefont {M.~W.}\ \bibnamefont {Mahoney}},\
  }\bibfield  {title} {\enquote {\bibinfo {title} {Community structure in large
  networks: {Natural} cluster sizes and the absence of large well-defined
  clusters},}\ }\href@noop {} {\bibfield  {journal} {\bibinfo  {journal}
  {Internet Math.}\ }\textbf {\bibinfo {volume} {6}},\ \bibinfo {pages}
  {29--123} (\bibinfo {year} {2009})}\BibitemShut {NoStop}%
\bibitem [{\citenamefont {Hall}, \citenamefont {Jaffe},\ and\ \citenamefont
  {Tratjenberg}(2001)}]{hall_nber_2001}%
  \BibitemOpen
  \bibfield  {author} {\bibinfo {author} {\bibfnamefont {B.~H.}\ \bibnamefont
  {Hall}}, \bibinfo {author} {\bibfnamefont {A.~B.}\ \bibnamefont {Jaffe}},\
  and\ \bibinfo {author} {\bibfnamefont {M.}~\bibnamefont {Tratjenberg}},\
  }\href@noop {} {\enquote {\bibinfo {title} {The {NBER} patent citation data
  file: {Lessons}, insights and methodological tools},}\ }\bibinfo {type}
  {Tech. Rep.}\ (\bibinfo  {institution} {National Bureau of Economic
  Research},\ \bibinfo {year} {2001})\BibitemShut {NoStop}%
\bibitem [{\citenamefont {Yang}\ \emph {et~al.}(2015)\citenamefont {Yang},
  \citenamefont {Zhang}, \citenamefont {Chen},\ and\ \citenamefont
  {Qu}}]{YANG2015170}%
  \BibitemOpen
  \bibfield  {author} {\bibinfo {author} {\bibfnamefont {D.}~\bibnamefont
  {Yang}}, \bibinfo {author} {\bibfnamefont {D.}~\bibnamefont {Zhang}},
  \bibinfo {author} {\bibfnamefont {L.}~\bibnamefont {Chen}},\ and\ \bibinfo
  {author} {\bibfnamefont {B.}~\bibnamefont {Qu}},\ }\bibfield  {title}
  {\enquote {\bibinfo {title} {Nationtelescope: Monitoring and visualizing
  large-scale collective behavior in {LBSNs}},}\ }\href@noop {} {\bibfield
  {journal} {\bibinfo  {journal} {Journal of Network and Computer
  Applications}\ }\textbf {\bibinfo {volume} {55}},\ \bibinfo {pages}
  {170--180} (\bibinfo {year} {2015})}\BibitemShut {NoStop}%
\bibitem [{\citenamefont {Backstrom}\ \emph {et~al.}(2006)\citenamefont
  {Backstrom}, \citenamefont {Huttenlocher}, \citenamefont {Kleinberg},\ and\
  \citenamefont {Lan}}]{BHKL06}%
  \BibitemOpen
  \bibfield  {author} {\bibinfo {author} {\bibfnamefont {L.}~\bibnamefont
  {Backstrom}}, \bibinfo {author} {\bibfnamefont {D.}~\bibnamefont
  {Huttenlocher}}, \bibinfo {author} {\bibfnamefont {J.}~\bibnamefont
  {Kleinberg}},\ and\ \bibinfo {author} {\bibfnamefont {X.}~\bibnamefont
  {Lan}},\ }\bibfield  {title} {\enquote {\bibinfo {title} {Group formation in
  large social networks: Membership, growth, and evolution},}\ }in\ \href@noop
  {} {\emph {\bibinfo {booktitle} {Proceedings of the ACM SIGKDD International
  Conference on Knowledge Discovery and Data Mining}}}\ (\bibinfo {address}
  {Philadelphia, PA, USA},\ \bibinfo {year} {2006})\ p.\ \bibinfo {pages}
  {44–54}\BibitemShut {NoStop}%
\bibitem [{\citenamefont {Kagan}, \citenamefont {Elovichi},\ and\ \citenamefont
  {Fire}(2018)}]{twitter}%
  \BibitemOpen
  \bibfield  {author} {\bibinfo {author} {\bibfnamefont {D.}~\bibnamefont
  {Kagan}}, \bibinfo {author} {\bibfnamefont {Y.}~\bibnamefont {Elovichi}},\
  and\ \bibinfo {author} {\bibfnamefont {M.}~\bibnamefont {Fire}},\ }\bibfield
  {title} {\enquote {\bibinfo {title} {Generic anomalous vertices detection
  utilizing a link prediction algorithm},}\ }\href@noop {} {\bibfield
  {journal} {\bibinfo  {journal} {Social Network Analysis and Mining}\ }\textbf
  {\bibinfo {volume} {8}},\ \bibinfo {pages} {27} (\bibinfo {year}
  {2018})}\BibitemShut {NoStop}%
\bibitem [{\citenamefont {Fire}\ and\ \citenamefont
  {Guestrin}(2020)}]{bitcoin}%
  \BibitemOpen
  \bibfield  {author} {\bibinfo {author} {\bibfnamefont {M.}~\bibnamefont
  {Fire}}\ and\ \bibinfo {author} {\bibfnamefont {C.}~\bibnamefont
  {Guestrin}},\ }\bibfield  {title} {\enquote {\bibinfo {title} {The rise and
  fall of network stars: Analyzing 2.5 million graphs to reveal how high-degree
  vertices emerge over time},}\ }\href
  {https://www.sciencedirect.com/science/article/pii/S0306457318308720}
  {\bibfield  {journal} {\bibinfo  {journal} {Information Processing \&
  Management}\ }\textbf {\bibinfo {volume} {57}},\ \bibinfo {pages} {102041}
  (\bibinfo {year} {2020})}\BibitemShut {NoStop}%
\bibitem [{\citenamefont {Stehlé}\ \emph {et~al.}(2011)\citenamefont
  {Stehlé}, \citenamefont {Voirin}, \citenamefont {Barrat}, \citenamefont
  {Cattuto}, \citenamefont {Isella}, \citenamefont {Pinton}, \citenamefont
  {Quaggiotto}, \citenamefont {Van~den Broeck}, \citenamefont {Régis},
  \citenamefont {Lina},\ and\ \citenamefont {Vanhems}}]{primary_school}%
  \BibitemOpen
  \bibfield  {author} {\bibinfo {author} {\bibfnamefont {J.}~\bibnamefont
  {Stehlé}}, \bibinfo {author} {\bibfnamefont {N.}~\bibnamefont {Voirin}},
  \bibinfo {author} {\bibfnamefont {A.}~\bibnamefont {Barrat}}, \bibinfo
  {author} {\bibfnamefont {C.}~\bibnamefont {Cattuto}}, \bibinfo {author}
  {\bibfnamefont {L.}~\bibnamefont {Isella}}, \bibinfo {author} {\bibfnamefont
  {J.-F.}\ \bibnamefont {Pinton}}, \bibinfo {author} {\bibfnamefont
  {M.}~\bibnamefont {Quaggiotto}}, \bibinfo {author} {\bibfnamefont
  {W.}~\bibnamefont {Van~den Broeck}}, \bibinfo {author} {\bibfnamefont
  {C.}~\bibnamefont {Régis}}, \bibinfo {author} {\bibfnamefont
  {B.}~\bibnamefont {Lina}},\ and\ \bibinfo {author} {\bibfnamefont
  {P.}~\bibnamefont {Vanhems}},\ }\bibfield  {title} {\enquote {\bibinfo
  {title} {High-resolution measurements of face-to-face contact patterns in a
  primary school},}\ }\href {https://doi.org/10.1371/journal.pone.0023176}
  {\bibfield  {journal} {\bibinfo  {journal} {PLoS ONE}\ }\textbf {\bibinfo
  {volume} {6}},\ \bibinfo {pages} {e23176} (\bibinfo {year}
  {2011})}\BibitemShut {NoStop}%
\bibitem [{\citenamefont {Zachary}(1977)}]{karate_club}%
  \BibitemOpen
  \bibfield  {author} {\bibinfo {author} {\bibfnamefont {W.~W.}\ \bibnamefont
  {Zachary}},\ }\bibfield  {title} {\enquote {\bibinfo {title} {An information
  flow model for conflict and fission in small groups},}\ }\href@noop {}
  {\bibfield  {journal} {\bibinfo  {journal} {Journal of Anthropological
  Research}\ }\textbf {\bibinfo {volume} {33}},\ \bibinfo {pages} {452--473}
  (\bibinfo {year} {1977})}\BibitemShut {NoStop}%
\bibitem [{\citenamefont {Lusseau}\ \emph {et~al.}(2003)\citenamefont
  {Lusseau}, \citenamefont {Schneider}, \citenamefont {Boisseau}, \citenamefont
  {Haase}, \citenamefont {Slooten},\ and\ \citenamefont
  {Dawson}}]{lusseau_bottlenose}%
  \BibitemOpen
  \bibfield  {author} {\bibinfo {author} {\bibfnamefont {D.}~\bibnamefont
  {Lusseau}}, \bibinfo {author} {\bibfnamefont {K.}~\bibnamefont {Schneider}},
  \bibinfo {author} {\bibfnamefont {O.~J.}\ \bibnamefont {Boisseau}}, \bibinfo
  {author} {\bibfnamefont {P.}~\bibnamefont {Haase}}, \bibinfo {author}
  {\bibfnamefont {E.}~\bibnamefont {Slooten}},\ and\ \bibinfo {author}
  {\bibfnamefont {S.~M.}\ \bibnamefont {Dawson}},\ }\bibfield  {title}
  {\enquote {\bibinfo {title} {The bottlenose dolphin community of {Doubtful}
  {Sound} features a large proportion of long-lasting associations. {Can}
  geographic isolation explain this unique trait?}}\ }\href@noop {} {\bibfield
  {journal} {\bibinfo  {journal} {Behav. Ecol. Sociobiol.}\ }\textbf {\bibinfo
  {volume} {54}},\ \bibinfo {pages} {396--405} (\bibinfo {year}
  {2003})}\BibitemShut {NoStop}%
\bibitem [{\citenamefont {Girvan}\ and\ \citenamefont
  {Newman}(2002)}]{girvan_community_2002}%
  \BibitemOpen
  \bibfield  {author} {\bibinfo {author} {\bibfnamefont {M.}~\bibnamefont
  {Girvan}}\ and\ \bibinfo {author} {\bibfnamefont {M.~E.~J.}\ \bibnamefont
  {Newman}},\ }\bibfield  {title} {\enquote {\bibinfo {title} {Community
  structure in social and biological networks},}\ }\href@noop {} {\bibfield
  {journal} {\bibinfo  {journal} {P. Natl. Acad. Sci. USA}\ }\textbf {\bibinfo
  {volume} {99}},\ \bibinfo {pages} {7821--7826} (\bibinfo {year}
  {2002})}\BibitemShut {NoStop}%
\end{thebibliography}%

\section*{Acknowledgements}

\noindent
This work has been supported in part by Slovenian Research Agency ARRS under the program P5-0168.
We gratefully acknowledge use of the Shark cluster of the LUMC for computation time.

\section*{Author contributions statement}

\noindent
V.T. conceived the experiments, V.T. and L.Š. designed the algorithms, 
V.T. and L.Š. analysed the results,
V.T. and L.Š. prepared the manuscript.
%V.T. wrote the manuscript.
Both authors reviewed the manuscript.

\section*{Data availability}

\noindent
The code generated during the current study is available in the \texttt{igraph} library, \url{https://github.com/vtraag/igraph/tree/flpa}.
The datasets analysed during the current study are available in the Netzschleuder repository, \url{https://networks.skewed.de}.

\section*{Competing interests}

\noindent
The authors declare no competing interests.

\end{document}